\documentclass[12pt]{iopart}

\usepackage{graphicx}
\usepackage[]{cite}

\begin{document}

\title[Structure and spatial distribution of Ge nanocrystals]
{Structure and spatial distribution of Ge nanocrystals subjected to fast neutron irradiation}

\author{S. Levy$^{* 1, 2}$, I. Shlimak$^{1, 2}$, D. H. Dressler$^{3}$, J. Grinblat$^{2}$, Y. Gofer$^{2, 3}$, T. Lu$^{4}$ and A. N. Ionov$^{5}$ }
\address{$^{1}$ Jack and Pearl Redneck Institute of Advanced Technology, Department of Physics, Bar-Ilan University, Ramat-Gan 52900, Israel }
\address{$^{2}$ Bar-Ilan Institute of Nanotechnology and Advanced Materials, Bar-Ilan University, Ramat-Gan 52900, Israel }
\address{$^{3}$ Department of Chemistry, Bar Ilan University, Ramat-Gan 52900, Israel }
\address{$^{4}$ Department of Physics and Key Laboratory for Radiation Physics Technology of Ministry of Education, Sichuan University, Chengdu 610064, P.R.China}
\address{$^{5}$ A. F. Ioffe Physico-Technical Institute RAS, St.-Petersburg 194021, Russia}
\ead{$^{*}$shai.levy10@gmail.com}

\begin{abstract}
The influence of fast neutron irradiation on the structure and spatial distribution of Ge nanocrystals (NC) embedded in an amorphous SiO$_2$ matrix has been studied. 
The investigation was conducted by means of laser Raman Scattering (RS), High Resolution Transmission Electron Microscopy (HR-TEM) and X-ray photoelectron spectroscopy (XPS).

The irradiation of NC-Ge samples by a high dose of fast neutrons 
lead to a partial destruction of the nanocrystals. 
Full reconstruction of crystallinity was achieved after annealing the radiation damage at 800$^o$C, which resulted in full restoration of the RS spectrum. HR-TEM images show, however, that the spatial distributions of NC-Ge changed as a result of irradiation and annealing. A sharp decrease in NC distribution towards the SiO$_2$ surface has been observed. This was accompanied by XPS detection of Ge oxides and elemental Ge within  both the surface and subsurface region. 
\end{abstract}

\submitto{\NT}

\maketitle

\section{Introduction}
Since the early 1990s, samples of Si and Ge nanocrystals (NC-Si, NC-Ge) embedded in a silicon dioxide (SiO$_2$) matrix have attracted much interest due to strong visible photoluminescence at room temperature \cite{Maeda,Kanemitsu}. Additional electrical studies performed on these NC have revealed charge retention properties \cite{3,4,5,6,7,8,9,10,11,12,13,shai}, suggesting the future incorporation of such structures in silicon-based electronic technology such as optoelectronic and microelectronic devices \cite{5,6,8,9,10,11,12,13,shai,15,16}. It has been shown \cite{17, 18} that NC-Ge's  smaller band-gap 
makes it a more suitable material than NC-Si for silicon-based technology. This property has made NC-Ge the promising candidate for the creation of a new generation of nano-scale optoelectronic devices. In this respect, investigation of the ability of the material to withstand different destructive influences is important for the design of semiconductor devices that can normally operate in a radiation-rich environment.

In the course of this analysis, the structure and spatial distribution of nanocrystals were investigated in NC-Ge samples which were irradiated in a research nuclear reactor with a high dose of neutrons, measuring several MeVs. Clash with fast neutrons causes displacement of atoms in the lattice in cascades, which produce a range of sub-nanometer clusters (defect complexes) and radiation damage \cite{Brinkman}. These directly contribute to the destruction of the optical and electrical performance of bulk Si and Ge-based devices (such as diodes and hetero-junction bipolar transistors (HBTs) \cite{20}). MOS (Metal Oxide Semiconductor) devices are known to withstand displacement-caused degradation only up to integral dose of 10$^{15}$ neutrons/cm$^2$ \cite{21}. Therefore, in this investigation, NC-Ge samples were subjected to much higher dose of $10^{20}$ neutrons/cm$^2$.

Measurements of Raman scattering (RS) and High Resolution Transmission Electron Microscopy (HR-TEM) are usually used to confirm the crystalline structure. 
Crystalline Ge is characterized by the Raman peak centered at approximately 300 cm$^{-1}$, which corresponds to the optical phonon frequency in bulk Ge crystals \cite{Fujii}. Measurements of HR-TEM images enable the examination of fine details of the NC structure. 
In addition to the crystalline structure analysis, the surfaces of the samples were analyzed by X-ray photoelectron spectroscopy (XPS).

\section{Experimental Details}
Our samples underwent five stages of treatments described below; at each stage of treatment a control sample was kept.

The NC-Ge samples were prepared using $^{74}Ge^+$ ion-implantation into a 500 nm thick amorphous SiO$_2$ layer, deposited on a silicon substrate with $\langle 100\rangle$ oriented surface. The details of the ion-implantation are documented in  \cite{Dun}. The Ge$^+$ ions were accelerated to 150 keV, with a dose of 1x10$^{17}$ ions/cm$^2$. These samples were labeled as ``implanted". A second treatment was conducted on the implanted samples by annealing at 800$^o$C. As a result, the randomly distributed Ge atoms formed nanocrystals (NC)\cite{Fujii2}. These samples were labeled as ``NC-Ge". The third process consisted of subjecting the NC-Ge samples to an intensive neutron irradiation in a research nuclear reactor, with the integral dose of $10^{20}$ neutrons/cm$^2$, and subsequently labeled as ``irradiated". After irradiation, a fourth  group consisting of irradiated NC-Ge samples was annealed at 600$^o$C to remove the radiation damage.  These samples were labeled as ``semifinal". The remaining samples were annealed a second time at 800$^o$C. This fifth group of samples was labeled as ``final".

RS spectra were measured with a Raman microscope LabRam HR at room-temperature using a 514.5 nm laser source for excitation. HR-TEM images were obtained using the 200 kV JEOL, JEM 2100 HR-TEM (LaB$_6$) integrated with a scanning device comprising annular dark-field and bright-field detectors and with a Noran System Six EDS (energy dispersive X-ray spectroscopy) system for elemental analysis. For cross-section imaging, HR-TEM samples were prepared by a focused ion beam (FIB) lift-out technique. XPS measurements were carried out on AXIS HS Kratos Analytical electron spectrometer system. The spectra were acquired using a monochromatic Al-K  (1486.6 eV)  X-ray source. 

\section{Results and discussion}
\subsection{Initial samples (before neutron irradiation) }
\begin{figure}[!h]
\center{
\includegraphics{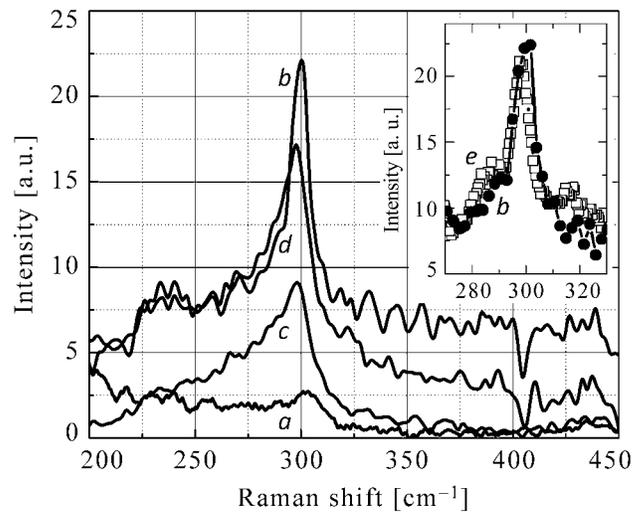}
\caption{Raman spectra of (\textit{a}) ``implanted", (\textit{b}) ``NC-Ge", (c) ``irradiated" and (\textit{d}) ``semifinal" samples; inset displays RS around 300 cm$^{-1}$ for ``NC-Ge" (\textit{b}) and ``final" (\textit{e}) samples.
\label{fig1}
}
}
\end{figure}

Curve \textit{a} in figure \ref {fig1} shows RS spectra in the ``implanted" sample, where Ge atoms do not yet form nanocrystals. One can see a broad asymmetric unstructured band which can be attributed to separate Ge atoms or amorphous nano-clusters \cite{Fujii2, 25}.
\vspace{1cm}
\begin{figure}[!h]
\center{
\includegraphics{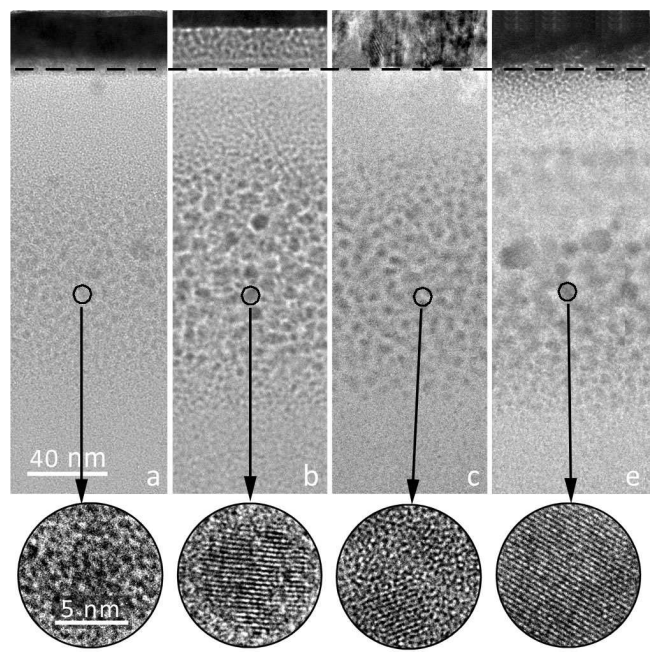}
\caption{HR-TEM cross-section of (\textit{a}) ``implanted" (\textit{b}) ``NC-Ge", (\textit{c}) ``irradiated" and (\textit{e}) ``final" samples. The dashed line indicates the surface of the samples covered by Pt which is needed for TEM cross-section preparation.
\label{fig2}
}
}
\end{figure}

Figure \ref{fig2}\textit{a} shows bright-field HR-TEM cross-sectional image of a Ge-implanted sample, where Ge atoms have not yet formed nanocrystals. In bright-field mode it is possible to distinguish between the Ge clusters and the SiO$_2$ matrix due to the contrast of the image, i.e. regions with Ge atoms appear darker while the SiO$_2$ matrix appears brighter. Nonappearance of NC is confirmed by the absence of a diffraction pattern.

Conforming to the SRIM simulation (Stopping and Range of Ions in Matter simulation) \cite{26}, the projection range and struggle (half-width of the distribution) for Ge ions with energy of 150 keV implanted in SiO$_2$ are approximately  100$\pm$5 nm and 30 nm, respectively, which concur with the experimental observation (figure \ref{fig2}\textit{a}).

In order to form nanocrystals, implanted samples were annealed at 800$^o$C. Formation of NC-Ge  was confirmed by the appearance of the main RS peak, centered at 300 cm$^{-1}$ (see curve \textit{b} in figure \ref{fig1}) which is characteristic of crystalline Ge. In addition, crystalline structure was observed in the enlarged HR-TEM image in the form of equidistant lattice planes (figure \ref{fig2}\textit{b}). The HR-TEM image shows that NC-Ge regions have an almost spherical shape with a diameter of approximately 2-10 nm. Inspection of the image reveals a correlation between the diameter and the depth of NC below the SiO$_2$ surface: with in the projection range (100 nm), where the density of the implanted Ge atoms is maximal, the average diameter of NC-Ge is also largest and symmetrically decreases on both sides.

\begin{figure}[!h]
\center{
\includegraphics{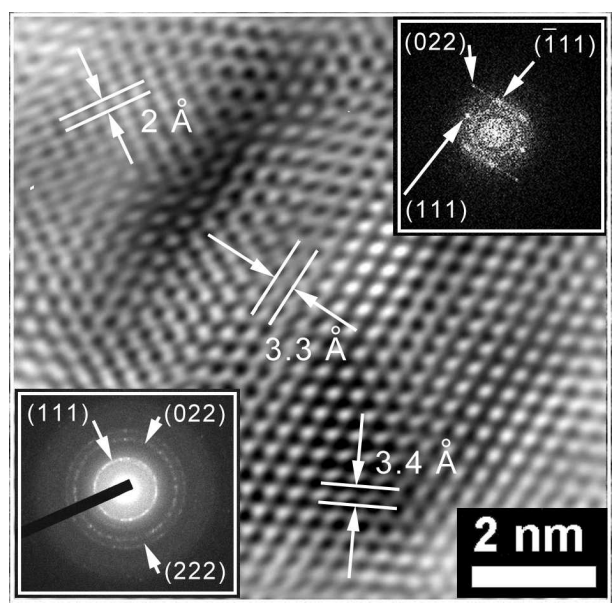}
\caption{Enlarged (processed) HR-TEM image of a single NC, live fast Fourier transform (FFT, upper right inset) calculated from the HR-TEM image and selected area electron diffraction (SAED, bottom left inset) of the NC-Ge layer inside the SiO$_2$ matrix. The d-spacing of the planes, as calculated by live FFT, are displayed in the HR-TEM image.
\label{fig3}
}
}
\end{figure}

The composition  of NC-Ge was studied by an accompanying investigation of  selected area electron diffraction (SAED) pattern of the NC-Ge layer inside the SiO$_2$ matrix and live fast Fourier transform (FFT) performed on an HR-TEM image of a single NC (figure \ref{fig3}). In SAED measurements, the d-spacing between planes (111), (022) and (222) was found to be 3.2, 1.95 and 1.67 {\AA} $\pm0.04$ {\AA}, respectively. In the case of FFT the d-spacing between planes (111), (022) and $(\bar{1}11)$ was found to be 3.3, 2.0 and 3.4 {\AA} $\pm0.06$ {\AA}, respectively.

Using both diffraction and live FFT, it is possible to conclude that the crystals consist of  Ge and are not a mix of any compound of Ge. It has also been observed that different crystallographic orientations coexist in some single NC, which can be explained by the existence of more than one nucleation center that serve as seeds forming each NC-Ge. That is, nucleation centers form small sized NC during the annealing process. These NC aggregate toward one another, and once they come into contact their lattice planes start to adjust at the interface. Nevertheless, the contact in the interface is not perfect.

\subsection{Influence of fast neutron irradiation and annealing}
Part of NC-Ge samples were irradiated by an intensive neutron flux (with energy up to a few MeVs) in a research nuclear reactor  with integral dose of 10$^{20}$ neutrons/cm$^2$.  As a result of this irradiation, a RS peak was decreased and broadened (curve \textit{c} in figure \ref{fig1}). This can be explained by the destruction of some NC due to the clash with fast neutrons and their subsequent transformation from Ge nanocrystals into amorphous clusters. However, RS peak did not disappear. This indicates that part of the NC-Ge has survived even after the high dose of destructive irradiation. The existence of remaining NC-Ge in irradiated samples was confirmed by HR-TEM image (figure \ref{fig2}\textit{c}). Taking into account that the dose of irradiation was five orders of magnitude larger than the one which destroyed the devices made from  bulk Ge and Si \cite{20}, one can conclude that this material shows potential for devices assigned to work in an extreme conditions, such as a radiation-rich environment.

In figure \ref{fig1}, curves \textit{d} and \textit{e} show the influence of the annealing of irradiated samples on the RS spectrum. For bulk Ge, disappearance of radiation damage induced by fast neutrons is achieved after annealing at 400-450$^o$C \cite{27}. For NC-Ge, however, annealing even at 600$^o$C leads only to a partial reconstruction of the RS peak (curve \textit{d} in figure \ref{fig1}); annealing at 800$^o$C is needed for full reconstruction of the initial RS peak (inset in figure \ref{fig1}). This phenomenon can be explained by the following: In irradiated bulk Ge, the destroyed areas are surrounded by a monocrystalline matrix that serves as a seed for recrystallization. In our samples, the destroyed areas are surrounded by amorphous SiO$_2$ which cannot promote the process of crystallization. As a result, the same temperature (800$^o$C) which was needed for the initial formation of NC-Ge after ion-implantation was still required in order to recover the crystalline structure after the fast neutron irradiation.

However, the reconstruction of crystallinity was not accompanied by the rebuilding of the initial space distribution of NC-Ge (compare figures \ref{fig2}\textit{b} and \ref{fig2}\textit{e}). It is evident that the spatial distribution of nanocrystals became asymmetric -- the increased average size still corresponds to the center of the projection range, while towards the SiO$_2$ surface, the density of NC falls more rapidly.

\begin{figure}[!h]
\center{
\includegraphics{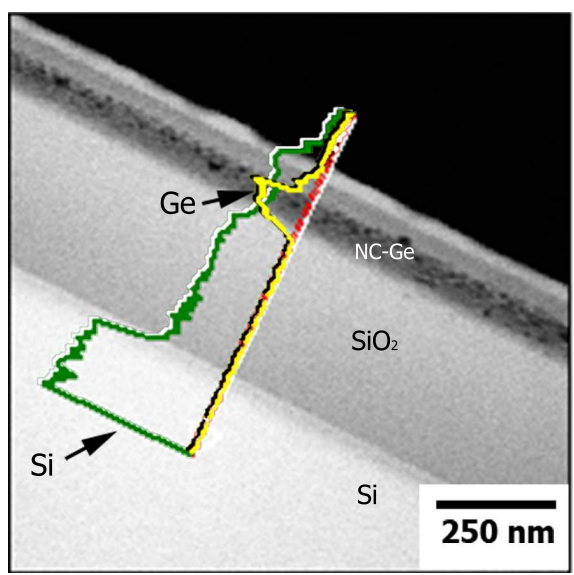}
\caption{STEM image including an EDS line scan profile of a ``final" sample. The curves display the abundance of elements with distance along the line (counts Vs. distance along the line). The curves  corresponding to elemental Ge and Si are shown.
\label{fig4}
}
}
\end{figure}

The asymmetry in the space distribution of NC-Ge after neutron irradiation and annealing was also confirmed by an EDS line scan of the final sample (figure \ref{fig4}). The EDS elemental mapping (line scan profile) provides information on the chemical composition of very small volumes of material by plotting the abundance of an element with distance along a line (counts Vs. distance). Figure \ref{fig4} shows the STEM image of the cross-section of a ``final" sample. The image clearly shows the asymmetry in distribution of Ge atoms inside the SiO$_2$ matrix.

\begin{figure}[!h]
\center{
\includegraphics {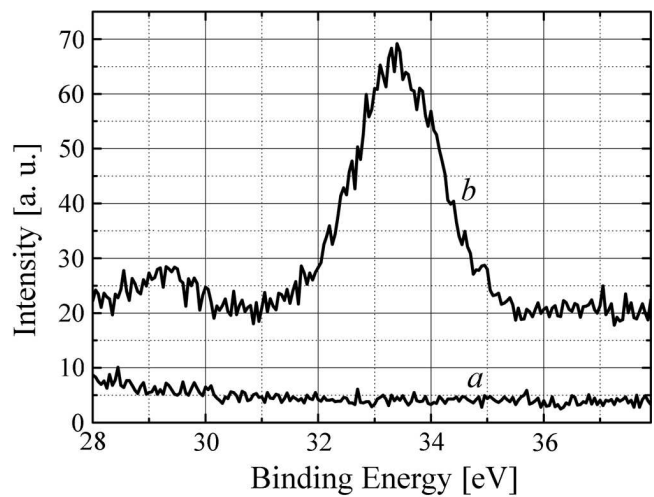}
\caption{High resolution X-ray photoelectron spectroscopy (XPS)  scans of Ge 3d peaks in ``NC-Ge"(\textit{a}) and ``final" (\textit{b}) samples.  The 33.4 eV peak attributed to Ge oxides (Ge 3d GeO$_2$), and a less pronounced peak at 29.4 eV  corresponding to elemental Ge are evident only in the``final" sample.
\label{fig5}
}
}
\end{figure}

This asymmetry can be explained by the influence of the near SiO$_2$ surface which attracts Ge atoms during the final annealing. XPS measurements were performed to confirm the assumption regarding the  enhanced diffusion of the Ge atoms towards the surface region  upon the annealing of the irradiated samples. The results are shown in figure \ref{fig5}. It is seen that in NC-Ge samples, the surface region is Ge free, while after neutron irradiation and a second annealing at 800$^o$C, a peak appears at 33.4 eV, attributed to Ge oxides (Ge 3d GeO$_2$), and a less pronounced peak at 29.4 eV  corresponds to elemental Ge \cite{Dutta}.

The presence of both Ge oxides and elemental Ge confirms that Ge atoms diffuse towards the surface during the second annealing from 600$^o$C to 800$^o$C.

\section{Conclusions}
Investigation of NC-Ge samples embedded in an amorphous SiO$_2$ matrix and subjected to irradiation by the neutron flux in a nuclear reactor demonstrates that part of the NC-Ge survives even after the exposure to a  high dose of irradiation. This makes this material promising for the fabrication of devices working in extreme conditions.
In the case of NC, annealing of radiation damage needs higher temperatures than that in bulk Ge. After the annealing of radiation damage, crystallinity is recovered, but the space distribution of NC-Ge becomes asymmetric due to the enhanced diffusion of Ge atoms towards the SiO$_2$ surface.

\ack
We are thankful to J. M. Lazebnik for his help in sample irradiation. We also thank M. Talianker,   A. Belostotsky and Y. Fleger for technical assistance. I. S. thanks the Erick and Sheila Samson Chair of Semiconductor Technology for financial support. This work was partly supported by the Israeli-China grant 3-405.

\end{document}